
\magnification=\magstep1
\baselineskip=24 true pt
\hsize=33 pc
\vsize=45 pc
\bigskip
\bigskip
\rightline {IP/BBSR/93-8}
\rightline {February, 1993}
\centerline {\bf TOPOLOGY CHANGING PROCESSES AND}
\centerline {\bf SYMMETRIES OF STRING EFFECTIVE ACTION}
\vfil
\centerline {{\bf S. Pratik Khastgir and Jnanadeva
Maharana}\footnote\dag{e-mail: pratik/maharana@iopb.ernet.in
}}
\bigskip
\centerline {\it { Institute of Physics, Bhubaneswar-751005, INDIA}}
\vfil
\centerline {\bf Abstract}

Wormhole solutions corresponding to space-time
geometries $R^1\times S^1\times S^2$ and $R^1\times S^3$ are obtained
from reduced string effective action and the action is written in a
manifestly $O(d,d)$ invariant form. A general treatment is given for
obtaining wormhole solutions of different topologies from dimensional
reduction. For specific ansatz of internal metric and antisymmetric
field the reduced action is shown to have a global $SL(2,C)$ symmetry.
The $SL(2,C)$ and duality symmetries have been exploited to generate
new configurations of internal fields which produce wormhole
solutions in four space-time dimensions. The $SL(2,C)$ symmetry
discussed in this paper arises due to specific form of the moduli and
these transformations belong to a subgroup of $O(d,d)$ global symmetry.
\vfil
\eject

\noindent {\bf I. Introduction:}

This paper is a sequel to our previous note [1] where new wormhole
solutions were obtained from the tree level string effective action.
It was shown that the effective action corresponding to the massless
modes of the string  admits wormhole solutions (in four Euclidean dimensions)
with topologies (geometries) $R^1\times S^3$ and $R^1\times S^1\times
S^2$. Further, it was shown how duality and $SL(2,R)$ transformations
can generate new wormhole solutions from a given classical
configuration. In the present paper we derive the solutions in more
general frame work with new explicit examples.

Let us recapitulate the essential steps to construct the string effective
action [2]. The evolution of the string, in the first quantized frame
work, is considered in the background of its massless modes. The conformal
invariance of the worldsheet action imposes stringent constraints
on the configuration of the background fields. These constraints
precisely correspond to the requirements that the $\beta$-functions
associated with the backgrounds must vanish in order to maintain the
the quantum conformal invariance of the theory. So long as one is
interested in the low energy effects of the string theory, it
suffices to consider the evolution of the string in the background of
its massless modes. The tree level string effective action, involving
only the massless excitations, can be constructed in such a way that
the equations of motion, derived from this effective action exactly
coincide with the vanishing of the $\beta$- functions.

In the recent past, considerable amount of effort has been devoted to the
study of the properties of the string effective action. The
investigations are focussed on two aspects: the classical solutions
of the equations of motion derived from
the string effective action and the underlying rich symmetry
structure of the action. Once a solution has been obtained we can generate new
background configurations by
implementing appropriate symmetry transformations. In what follows, we
first discuss the symmetries and then the classical solutions.

It is recognized that one of the important symmetries in string theory
is the target
space duality [3]. For example, when one considers a compactified string
on a circle of radius $R$, the spectrum of the string remains
invariant under the transformation $R\rightarrow {\lambda^2\over R}
({\lambda^2=2\alpha'\hbar})$. The consequences of duality have been
explored in various directions. It has led to the introduction of a
minimum compactification scale and has been employed to restrict
the form of scalar (or super) potential leading to the study of the
nonperturbative mechanisms of
supersymmetry breaking. Moreover, the consequences of duality in
cosmology [4] are also very interesting.

Next, the global $O(d,d)$ symmetry of the string effective action,
discovered by Meissner and Veneziano [5,6] has attracted considerable
amount of attention in recent years, where the massless modes are
only allowed to be time dependent. It was shown that the effective
action is invariant under the global $O(d,d)$ transformations where
$d$ is the number of spatial dimensions and the space-time dimension
$D=d+1$. Sen and collaborators [7] extended above results when
backgrounds are independent of $d$ coordinates (dependent on $D-d$
coordinates) and showed that one can perform a global $O(d)\times
O(d)$ transformation on the solution to generate a new background
configuration satisfying the classical equations of
motion to all orders in the string tension $\alpha'$. Later, similar results
were also obtained by them for heterotic strings. Recently,
Maharana and Schwarz [8] have shown the invariance of tree level string
effective action under $O(d,d)$ transformations by employing dimensional
reduction. Authors started with the $D+d$
dimensional string effective action, where $d$ is the number of
compact coordinates and obtained the $D$ dimensional reduced action.
They have shown that the reduced action is manifestly invariant under
the global $O(d,d)$ transformation. The technique was extended to the
case when there are $n$ Abelian vector fields in $D+d$ dimensions.  The $D$
dimensional reduced action has $O(d,d+n)$ symmetry (earlier studied
by Hassan and Sen for heterotic strings) provided that the
Chern-Simons term is included in
the antisymmetric tensor field strength for the bosonic string. The
reduced action can be
reformulated on suitable coset spaces with global and local
symmetries. An interesting application is to generate  non trivial
geometries by implementing the $O(d,d)$ transformations on a trivial
background (flat geometries). In other words, these $O(d,d)$ transformations
relate different string vacua (geometries) which are not equivalent in
general.

The four dimensional string effective action possesses another
important global symmetry. Let us recall that the equations of motion
of pure electromagnetic field coupled to gravity are invariant under
duality transformations such that electric and magnetic fields are
continuously rotated into one another. However, when a dilaton field is
coupled to the above system the duality transformation is discrete.
Recently, Shapere $et\;al.$ [9,10] have shown that, with the introduction of
an axion field, the duality symmetry can be made continuous. It is well
known that axion field appears naturally in the low energy string effective
action since, in four dimensions, it is dual to the antisymmetric field
strength. Furthermore, the axion and the dilaton belonged to the same
supermultiplet(in supergravity theories) and in this generalized duality
the axion and the
dilaton are treated as the real and imaginary part of a complex field
respectively. Indeed the multiplet transforms as an $SL(2,R)$ vector.
The dyon solution was constructed from the pure magnetic one using the
above $SL(2,R)$ transformation. It is important to note that the
effective action is not invariant under the $SL(2,R)$ transformations
whereas the equations of motion derived from the action are.

Recently, several coset models [11] have been constructed which provide
examples of exact conformal field theories for the string
backgrounds. These are gauged WZNW
models where one starts with an ungauged WZNW model with group
manifold G. Subsequently, a subgroup H is gauged such that the
underlying conformal field theory is shown to be exactly solvable and the
backgrounds satisfy the $\beta$- function equations. Thus the choice
of group manifold amounts to a given background which is to be
consistent with conformal invariance. Connection between the WZNW model
and the $O(d,d)$ transformation was established by Hassan and Sen
[11]. Their analysis makes a general conjecture that $O(d,d)$
transformations on any WZNW model correspond to marginal deformations
of the WZNW theory by appropriate combination of left and right
moving currents belonging to the Cartan subalgebra.

The classical solutions satisfying the $\beta$-function equations can
be broadly classified as follows: i) Cosmological ii) Black hole(string/brane),
and iii) Wormhole type.

The non trivial conformal backgrounds of cosmological [12] interest were
generated from the trivial space-time metrics. These were mainly
obtained by `boosting' the flat Milne or Rindler metrics. Here by boosting
we mean implementing special $O(d,d)$ transformation on trivial backgrounds.

Black hole solutions attracted enormous attention in last couple of
years. Initially the four dimensional black hole solutions were
obtained in Refs. [13]. More recently, Witten [11] obtained an exact conformal
field theory describing a two dimensional stringy black hole as an
$SL(2,R)/U(1)$ WZNW model. Several other stringy black hole solutions [14]
analogous to Reissner-Nordstrom, Kerr etc. type
were obtained in two and higher dimensions.

The last in our listing is solutions of the wormhole type, which
is our topic of main interest. Wormholes received considerable attention when
Coleman [15] suggested that they may solve the problem of cosmological
constant persuing the arguments of the Baum and Hawking [16].
Coleman has advocated that the wormholes introduce quantum
indeterminacy in the constants of nature [17]. Thus, it has been proposed
that the fundamental constants are endowed with a probability
distribution function for universes and it is necessary to take the
ensemble average over all the universes. Earlier, it was noted by
Baum and Hawking [16] that the Euclidean partition function is highly
peaked around $\Lambda=0$. It has been argued in Ref. [17] that the sum
over wormhole topology shifts effective coupling constants, making
$\Lambda$ an integration variable with distribution peaked at
vanishing of cosmological constant. The effects of these topology
changing processes are really surprising [18,19]. Hawking has, on the
other hand, argued the  possibility of
effective loss of quantum coherence [20] in quantum gravity due to
the wormholes. When the
metric is treated as a quantum field, the topology of space-time may
fluctuate on the Planck scale. Planck sized baby universes might be
formed. These baby universes can carry away information with them
resulting in an effective loss of quantum coherence. Coleman again
argued that quantum incoherence will not be observed if the many
universe system is in an equilibrium state because the rate of
coherence loss depends on the state of the many universe system. Thus
the cosequences of the topology changing processes have attracted
considerable amount of attention in the recent past and have been
explored extensively [21].
 If the wormhole corresponds to a saddle point of the Euclidean
action, then the semiclassical approximations can be employed in
order to facilitate computations of vertex operators and correlation
functions in the path integral formalism [22].
Indeed, axionic wormhole solutions have been found by Giddings and
Strominger [19] in four dimensions
for effective actions that arise in string theories. These
actions are to be envisaged as derived from tree level effective
action of bosonic/heterotic string theories in critical dimensions
where the internal dimensions are compactified on suitable geometries.

The main objective of this investigation is to get wormhole
solutions from string effective action.
The paper is organized as follows: In the section II, we present known
wormhole solutions obtained from an action where antisymmetric tensor
field is coupled to gravity. The first solution is the well known
Giddings-Strominger wormhole [19] having topology $R^1\times S^3$. The
second one is having a topology $R^1\times S^1\times S^2$ and can be
transformed to the one obtained by Keay and Laflamme [23]
by a coordinate transformation. In section III, the dimensional reduction
technique of Maharana and Schwarz [8] is summarized. The $O(d,d)$
symmetry of the dimensionally reduced action is described in brief.
Finally equations of motion for the various fields are obtained.
In section IV, we obtain the wormhole solutions from the effective action. We
explicitly show how $R^1\times S^3$ and $R^1\times S^1\times S^2$
geometries can be obtained from string compactification. We present
explicit forms of the various backgrounds fields. Then we discuss the
symmetries of the reduced action and show how to generate new solutions.
Here, it is shown that the reduced action is invariant under a global
$SL(2,C)$ symmetry; exploiting this symmetry, we generate new solutions.
It was further shown that the $SL(2,C)$ symmetry is a part of the bigger
global $O(d,d)$ symmetry.
A general treatment of wormholes from the strings is presented in the
section V. We show here how different topologies can be obtained
from compactification or dimensional reduction. Finally
in the section VI summary and discussions are given.

\noindent {\bf II. Antisymmetric tensor field coupled to gravity:}

We first present two wormhole solutions of different topologies
corresponding to the same action. One of these is the wellknown
``Giddings-Strominger wormhole" [19]  with the topology $R^1\times S^3$.
We obtain another solution with the geometry $R^1\times S^1\times
S^2$. We start with the same action considered by Giddings and
Strominger in four dimensions. This is an action for pure gravity
coupled to the antisymmetric field tensor. The boundary term is
suppressed here. Note that the presence of the boundary term does not
affect the equations of motion and wormhole solutions. The four
dimensional Euclidean action reads,

$$S_4 = \int d^4x \sqrt {g}~\bigg\{- R+{1\over 12}H_{\mu \nu
\rho} ~H^{\mu \nu \rho}\bigg\}. \eqno (1)$$

\noindent where, three-form $H=dB$.
We have set gravitational and other couplings equal to one.

i) {\bf $u
bf R^1\times S^3}$ Geometry:} Now we discuss the
Giddings-Strominger wormhole solution in brief. The space has the topology
$R^1\times S^3$ for this solution. To start with,
the following ansatz is chosen for the 4-dimensional
metric and the antisymmetric tensor field strength,

$$ds^2=dr^2+R^2(r)d\Omega_3^2,\qquad H=2\sqrt {3}
R_0^2 \sin^2\theta \sin\psi~ d\theta\wedge d\psi
\wedge d\phi. \eqno(2)$$

\noindent $d\Omega^2_3$ is the line element on the three sphere.
$R_0$ appearing in the field strength is a constant.
In this case (rr) Einstein equation becomes,

$$R'^2=1-{R_0^4\over R^4}.\eqno (3)$$

\noindent Here and everywhere the prime denotes derivative with respect to
$r$. All other equations of motion (the remaining Einstein
equations and the $H$ equations) are satisfied if one uses the ansatz
(2) with the relation (3). The matter field equation involving
$H_{\mu\nu\lambda}$ is only a charge conservation equation.
$R(r)$ appearing in the metric is the scale
factor. The detailed discussions of solution and other
properties can be found in Ref. [19]. We mention in passing that the
scale factor $R^2(r)\rightarrow r^2$ for
$r\rightarrow \pm\infty$, corresponds to the two asymptotically
flat regions. Here the radius $R(r)$ of $S^3$ contracts from infinite
value to a minimum neck size $R_0$ and again expands to infinity.
The line element covers only half the wormhole, from
asymptotically flat region to the throat.
The Euclidean action of the ``semiwormhole" [19] is,

$$ S =2\int d^4x {\sqrt g} {1\over 12} H^2$$
$$~~~=6\pi^3R_0^2. \eqno(4)$$

\noindent Half of the $S$ comes from the scalar curvature appearing
in the action (1)
and the half from the antisymmetric field strength.
The volume integral of $H$ is the axionic charge, $Q$, proportional
to the throat size $R_0$.

ii) {\bf ${\bf R^1\times S^1\times S^2}$ Topology:} The wormhole solution with
$R^1\times R^1\times S^2$ geometry was first discussed by Morris and
Thorne in Ref [24]. In their solution, no explicit matter fields  were
mentioned which could give rise to required energy-momentum tensor.
The metric was Lorentzian in nature. Later, Keay and Laflamme [23]
obtained an Euclidean wormhole solution with topology
 $R^1\times S^1\times S^2$ for gravity coupled to antisymmetric
tensor field. We first present here the $R^1\times S^1\times S^2$
geometry solution in an elegant form so that it can be compared with
Giddings-Strominger solution easily. Next we give the transformation
which will take our form to the Keay and Laflamme solution. Effective
action is same as (1).
The ansatz for line element and antisymmetric field is chosen as,

$$ds^2=R_1^2dt^2+dr^2+R^2(r)d\Omega_2^2\qquad H=2R_1R_0
\sin\theta~ dt\wedge d\theta\wedge d\phi. \eqno(5)$$

\noindent Here, $d\Omega_2^2$ is the line element on the two sphere.
The coordinate $t$ has a range $[0,2\pi]$. Using above
ansatz in the (rr) Einstein equation one gets,

$$R'^2=1-{R_0^2\over R^2}.\eqno (6)$$

\noindent Again ansatz (5) together with the relation (6) satisfy all
equations of motion. The solution of the equation (6) is as follows:

$$R^2=R^2_0+r^2. \eqno (7)$$

\noindent Here $R_0$ is a constant. Solution, $R(r)\rightarrow \pm r$
as $r\rightarrow \pm \infty$, corresponds to the two
asymptotically flat Euclidean regions. This wormhole solution
corresponds to a fixed radius $R_1$ of the circle $S^1$. The
radius $R(r)$ of the sphere $S^2$ contracts from an infinitely
large value to a minimum throat size $R_0$, and reexpands to
infinity. Now to transform this to the form given in Ref. [22] one has
to use, $r^2\rightarrow l^2-R^2_0$, so that, $dr^2\rightarrow
(1-{R^2_0\over l^2})^{-1}dl^2$. The ``semiwormhole" action,

$$ S =2\int d^4x {\sqrt g} {1\over 12} H^2=
2\int_0^{2\pi}dt\int_0^{\infty}dr\int_0^{\pi}d\theta\int_0^{2\pi}d\phi
{\sqrt g}~ 2R_1\sin \theta {R_0^2\over R^2}= 16 \pi^3 R_0R_1. \eqno (8)$$

\noindent  Notice that if `$t$' is identified as a
compact coordinate the $B_{t\phi}$ component has the interpretation
of a three dimensional gauge field such that
${\bf A}=(0,0,2R_0R_1(1-\cos \theta))$.
The corresponding field strength $F^2=8R_0^2R_1^2/R^4$ is analogous to
that of a magnetic monopole with above ansatz (5) for $H$. Thus we
recover the $2+1$ dimensional wormhole solution of Gupta $et\; al.$ [25].

\noindent {\bf III. Dimensional reduction, $O(d,d)$ symmetry, and equations
of motion:}

In Ref. [8] authors employed Scherk-Schwarz [26] dimensional
reduction technique in order to derive string effective action in
lower dimensions and to investigate the origin of the noncompact
symmetries in string theory. Let us begin by recalling  the main results
of Ref. [8] and set the notations. The
bosonic part of the effective action in
 $\hat D=D+d$ Euclidean dimensions ($\hat D=26,10$ for bosonic,
fermionic critical cases respectively) is,

$$S_{\hat g} = \int d^{\hat D}x~ \sqrt{ \hat g}~ e^{-\hat\phi}
\big [-\hat R
(\hat g) - \hat g^{\hat \mu \hat \nu} \partial_{\hat \mu} \hat\phi
\partial_{\hat \nu} \hat\phi + {1 \over 12} ~ \hat H_{\hat \mu \hat \nu
\hat \rho} ~
\hat H^{\hat \mu \hat \nu \hat \rho}\big ].\eqno (9)$$

\noindent $\hat H$ is the field strength of antisymmetric tensor and $\hat
\phi$ is the dilaton. Here we have set all the nonabelian gauge field
backgrounds to zero. When the backgrounds are independent of the `internal'
coordinates $y^{\alpha}, \alpha=1,2..d$ and the internal space is
taken to be torus, the metric $\hat g_{\hat \mu \hat \nu} $
 can be decomposed as

$$\hat g_{\hat \mu \hat \nu} = \left (\matrix {g_{\mu \nu} +
A^{(1)\gamma}_{\mu} A^{(1)}_{\nu \gamma} &  A^{(1)}_{\mu \beta}\cr
A^{(1)}_{\nu \alpha} & G_{\alpha \beta}\cr}\right ),\eqno (10)$$

\noindent where $G_{\alpha \beta}$ is the internal metric and $g_{\mu\nu}$,
the $D$-dimensional space-time metric, depend on the coordinates $x^{\mu}$.
The dimensionally reduced action is,

$$\eqalign {S_{\hat g} =& \int d^Dx \sqrt {g}~ e^{-\phi}
\bigg\{- R - g^{\mu \nu}
\partial_{\mu} \phi \partial_{\nu} \phi +{1\over 12}H_{\mu \nu \rho} ~
H^{\mu \nu \rho}\cr
&- {1 \over 8} {\rm tr} (\partial_\mu M^{-1} \partial^\mu
M)+ {1 \over 4}
{\cal F}^i_{\mu \nu} (M^{-1})_{ij} {\cal F}^{\mu \nu j} \bigg\}}.
\eqno (11)$$

\noindent Here $\phi=\hat\phi-{1\over 2}\log\det G$ is the shifted dilaton.
$$H_{\mu \nu \rho} = \partial_\mu B_{\nu \rho} - {1 \over 2} {\cal A}^i_\mu
\eta_{ij} {\cal F}^j_{\nu \rho} + ({\rm cyc.~ perms.}),\eqno (12)$$
\noindent ${\cal F}^i_{\mu \nu}$ is the $2d$-component vector of field
strengths
$${\cal F}^i_{\mu \nu} = \pmatrix {F^{(1) \alpha}_{\mu \nu}
\cr F^{(2)}_{\mu \nu \alpha}\cr} = \partial_\mu {\cal A}^i_\nu - \partial_\nu
{\cal A}^i_\mu \,\, ,\eqno (13)$$
\noindent $A^{(2)}_{\mu \alpha} = \hat B_{\mu \alpha} + B_{\alpha \beta}
A^{(1) \beta}_{\mu}$ (recall $B_{\alpha \beta}=\hat B_{\alpha \beta}$), and
the $2d\times 2d$ matrices are
$$M = \pmatrix {G^{-1} & -G^{-1} B\cr
BG^{-1} & G - BG^{-1} B\cr},\qquad \eta =  \pmatrix {0 & 1\cr 1 & 0\cr}
\, .\eqno (14)$$

\noindent The action (3) is invariant under a global $O(d,d)$ transformation,

$$M \rightarrow \Omega M \Omega^T, \qquad \Omega^T \eta \Omega = \eta, \qquad
{\cal A}_{\mu}^i \rightarrow \Omega^i{}_j {\cal A}^j_\mu, \qquad {\rm
where} \qquad
\Omega \in O(d,d). \eqno (15)$$
\noindent Note that $M\in O(d,d)$ also and $M^T\eta M=\eta$. The
background equations of motion can be derived from (11). The classical
solutions of string effective action correspond to different string vacua
and are given by solutions for $M$,$\cal F$ and $\phi$.

The equations of motion derived from the reduced action (11),
the Einstein equation and matter field equation, are as
follows:

\noindent Einstein-equation:
$$R_{\mu\nu} +\partial_{\mu} \phi \partial_{\nu} \phi
-{1\over 12}H_{\mu \lambda\rho}~H_{\nu}^{~\lambda \rho}
+ {1 \over 8} {\rm tr} (\partial_\mu M^{-1} \partial_\nu
M)- {1 \over 2}
{\cal F}^i_{\mu \lambda} (M^{-1})_{ij} {\cal F}_{\nu}^{~\lambda j} $$
$$ -{1\over 2} g_{\mu\nu}\big [ R + g^{\lambda \rho}
\partial_{\lambda} \phi \partial_{\rho} \phi
-{1\over 12}H_{\lambda \rho \sigma} ~
H^{\lambda \rho \sigma}
+ {1 \over 8} {\rm tr} (\partial_\lambda M^{-1} \partial^\lambda
M)- {1 \over 4}
{\cal F}^i_{\lambda \rho} (M^{-1})_{ij}
{\cal F}^{\lambda \rho j}\big ]=0, \eqno(16)$$

\noindent Dilaton-equation:
$$2\partial_{\mu}({\sqrt g} e^{-\phi}g^{\mu\nu}\partial_{\nu}\phi)
-{\sqrt g}e^{-\phi}\big [- R - g^{\mu \nu}
\partial_{\mu} \phi \partial_{\nu} \phi +{1\over 12}H_{\mu \nu \rho} ~
H^{\mu \nu \rho} $$
$$ - {1 \over 8} {\rm tr} (\partial_\mu M^{-1} \partial^\mu
M)+ {1 \over 4}
{\cal F}^i_{\mu \nu} (M^{-1})_{ij} {\cal F}^{\mu \nu j}\big ]=0, \eqno(17)$$

\noindent$B_{\mu\nu}$-equation:
$$\partial_{\lambda}({\sqrt g} e^{-\phi}H^{\lambda\mu\nu})=0, \eqno(18)$$

\noindent Equation for gauge fields:
$${1\over 4}{\sqrt g}e^{-\phi}\eta_{ki}{\cal
F}_{\nu\rho}^iH^{\mu\nu\rho} -{1\over 2}\partial_{\lambda}\big [{\sqrt
g}e^{-\phi} \{{\cal
A}_{\nu}^i\eta_{ik}H^{\nu\lambda\mu}+2(M^{-1})_{ki}{\cal
F}^{\lambda\mu i}\}\big ]=0, \eqno(19)$$

\noindent Equation of motion for $M$ is a matrix equation.
$$\partial_\mu ({\sqrt g} e^{-\phi} g^{\mu\nu}M_{kl}\eta_{lm} \partial_\nu
M_{mi})+{\sqrt g} e^{-\phi}{\cal F}_{\mu\nu i}M_{kl}\eta_{lm}{\cal
F}^{\mu\nu}_{m} -{\sqrt g} e^{-\phi}{\cal
F}_{\mu\nu m}\eta_{ml} M_{li}{\cal F}^{\mu\nu}_{k}=0. \eqno(20)$$

\noindent Let us briefly discuss about
the notations and content of the last equation.
The lower case Latin subscripts specify the elements of the
matrices and the vectors appearing in the equations. ${\cal
F}_{\mu\nu}^i \equiv {\cal
F}_{\mu\nu i}$ and ${\cal
F}^{\mu\nu i} \equiv {\cal
F}^{\mu\nu}_ i$ (this type of notation for the $O(d,d)$ indices were
used all along in this section). The equation (20)
should be understood in following sense to be consistent with our
earlier notation. It has been already mentioned that the matrix $M$
has $2d\times 2d$ elements. The upper diagonal $d\times d$ block
consisting $G^{-1}$ denoted by two upper indices whereas the lower
diagonal block elements are denoted by two lower indices; the off-diagonal
blocks
are written with one lower and one upper index.

\noindent {\bf IV. Wormholes from strings:}

Now we would like to obtain the
wormhole solutions of different topologies in four dimensional
spacetime starting from a suitable string effective action.

\noindent {\bf IV.1. Field equations and solutions:}

\noindent i) {${\bf R^1\times S^3}$:} In this subsection we shall
obtain $R^1\times S^3$
wormhole solution from heterotic string in critical dimensios. Here
we have $\hat D=10=4+6$, six being the number of `internal' dimensions. The
backgrounds $\phi$ and $B_{\mu\nu}$  are chosen to be
constants and the background ${\cal A}_{\mu}^i$=0. We choose spherically
symmetric ansatz for internal dimensions.
 The internal $6\times 6$
matrix, $(G+B)_{\alpha\beta}, \alpha,\beta=1,..6$, is represented as
three blocks of $2\times 2$ matrices of the following form.

$$G+B = \pmatrix {\Sigma_1 & 0 & 0 \cr 0 & \Sigma_2 & 0 \cr
0 & 0 & \Sigma_3 \cr},\qquad
\Sigma_j= \pmatrix {e^{\lambda_jD_j(r)} &
a_j(r)\cr -a_j(r) & e^{\lambda_jD_j(r)}\cr}, \eqno (21)$$

\noindent no summation over the repeated index $j$ is understood
above. The effective action (11)
takes the form

$$S_4 = \int d^4x \sqrt {g}~\bigg [- R+{1\over 2}g^{rr} \sum_{j=1}^3
\bigg \{\lambda_j^2
(\partial_rD_j)^2 +e^{-2\lambda_jD_j}(\partial_ra_j)^2\bigg \}\bigg ].
\eqno (22)$$

\noindent The equations of motion (16-20) for this ansatz reduce to,

$$-R+{1\over 2}g^{rr}\sum_{j=1}^3\bigg \{\lambda_j^2
(\partial_rD_j)^2 +e^{-2\lambda_jD_j}(\partial_ra_j)^2 \bigg \}=0,\eqno(23)$$

$$R_{rr}-{1\over 2}g_{rr}R={1\over 4}\sum_{j=1}^3\bigg \{\lambda_j^2
(\partial_rD_j)^2 +e^{-2\lambda_jD_j}(\partial_ra_j)^2 \bigg \},\eqno (24)$$

$$R_{ii}-{1\over 2}g_{ii}R=-{1\over 4}g_{ii}g^{rr}\sum_{j=1}^3
\bigg \{\lambda_j^2 (\partial_rD_j)^2
+e^{-2\lambda_jD_j}(\partial_ra_j)^2\bigg \}
,\qquad i\neq r,\eqno (25)$$

$$\partial_r({\sqrt g}g^{rr}e^{-2\lambda_j D_j}\partial_r a_j)=0,\eqno(26)$$
\noindent and
$$\partial_r({\sqrt g}g^{rr}\lambda_j\partial_r D_j)
+{\sqrt g}g^{rr}e^{-2\lambda_jD_j}(\partial_r a_j)^2=0.\eqno(27)$$

\noindent We mention in passing that equations (26) and (27) hold for
each $j$=1,2 and 3; moreover these equations represent the axionic charge
conservation and the radial dilaton evolution respectively. Whereas
eqns. (24) and (25) are Einstein equations. Equation (23) is the
reduced equation
obtained from equation (17). We would have lost this equation had we
deived the equations from the reduced action. Notice that the (23)
does not provide any new information. One can obtain this equation
from equations (24) and (25) using the ansatz we have chosen for the
metric. The remaining equations
(18) and (19)
(the antisymmetric field tensor $B_{\mu\nu}$ and the gauge field
${\cal A}_{\mu}^i$ equations respectively)
are trivially satisfied for the ansatz used for the backgrounds in
the beginning. The choice,

$$ds^2=dr^2+R^2(r)d\Omega_3^2,\eqno(28a)$$
$$e^{-2\lambda_jD_j}={Q_j^2\over R^4},\qquad a_j=\pm{i\over Q_j}R^2R,'
\eqno(28b)$$

\noindent with the relation $R'^2=1-{R_0^4\over R^4}$, R(r) being the
radius of $S^3$ with a throat size $R_0$, mention equations (23)-(27).
We observe that both the fields, radial dilaton and axion, no where
become singular for finite $r$, but diverges for $r\rightarrow \pm \infty$.
It is worthwhile to point out that the axionic background envisaged
above has similarity with the axion solution adopted by Giddings
and Strominger [27]. However, our backgrounds arise from the
prescriptions of toroidal compactifications [8]. On the other hand,in
Ref. [27], the antisymmetric tensor field, $B_{\alpha\beta}$,
associated with the internal dimensions,
are the fundamental K{\"a}hler forms of the internal dimensions and
their $x$ and $y$ dependence are decomposed in a specific manner.
Moreover, we have a special choice of $G$ and $B$ and have
obtained explicit form for the radial axion field.

\noindent ii) {${\bf R^1\times S^1\times S^2}$:} We start with a
 closed bosonic string
in critical dimensions ($\hat D=D+d=4+22=26$). The configuration is
such that the fourteen out of its twentytwo `internal' dimensions are flat.
Next we choose the backgrounds $\phi$=const., $B_{\mu\nu}$=const. and
${\cal A}_{\mu}^i$=0.
The remaining eight `internal' $(G+B)_{\alpha\beta} (\alpha,\beta=1,..8)$
matrix is decomposed in a fashion similar to the previous case. In
this case we have only four blocks.
$$G+B = \pmatrix {\Sigma_1
& 0 & 0 & 0\cr 0 & \Sigma_2 & 0 & 0\cr
0 & 0 & \Sigma_3 & 0\cr 0 & 0 & 0 & \Sigma_4\cr}, \eqno (29)$$
\noindent where $\Sigma_j$ is again given by (21). The action and the
equations of motion look exactly same as the action (22) and the
equations of motion (23)-(27) except the fact that now the index `$j$'
runs from 1 to 4 only unlike the previous case where it was 1 to 3.
The solution obtained is as follows:

$$ds^2=R_1^2dt^2 + dr^2+R^2(r)d\Omega_2^2,\eqno(30a)$$
$$e^{-2\lambda_jD_j}={Q_j^2\over R^2},\qquad a_j=\pm{i\over Q_j}RR',
\eqno(30b)$$
\noindent with the relation $R'^2=1-{R_0^2\over R^2}$, satisfies all
the equations of motion.
Where $R(r)$ is the scale factor and $R_0$ gives size of the wormhole
neck. Again we see that the solution does not become singular
anywhere for finite `$r$'.
So we get the desired $R^1\times S^1\times S^2$ from the
critical bosonic string. The solutions presented in eqn. (30) are the
simplest ones for the geometry $R^1\times S^1\times S^2$; there might
be other solutions for the same geometry.

\noindent {\bf IV.2. Symmetries:}

We discuss below the symmetries of the reduced effective action and
generate new solutions by implementing the symmetry transformations.

\noindent i) {\bf ${\bf O(d,d)}$ Symmetry:} It was already
shown that the action (11) is invariant under
global $O(d,d)$ transformation for space-time dependent $G$ and $B$.
Hence the reduced action (22) is also manifestly $O(6,6)$ invariant.
The  duality
transformations $M\rightarrow\ \eta M \eta =M^{-1}$ is a special form
of the above global noncompact transformations. The new
backgrounds (duality transformed) ${\tilde a_j}$ and ${\tilde D_j}$ are
given by
$$e^{-2\lambda_j\tilde {D}_j}=(e^{\lambda_jD_j}+e^{-\lambda_jD_j}a_j^2)^2=
\bigg ({{R_0^4}\over {Q_j^2}}\bigg )^2e^{-2\lambda_jD_j},\eqno (31a)$$

\noindent and
$$\tilde {a}_j=-(e^{\lambda_jD_j}+e^{-\lambda_jD_j}a_j^2)^{-1}
e^{-\lambda_jD_j}a_j=-{{Q_j^2}\over {R_0^4}}a_j.\eqno (31b)$$
\noindent It is interesting to note that for $Q_j=R_0^2$ the axionic
charges of the original theory and the transformed theory are the
same. Thus such a value of $Q_j$ corresponds to a self dual theory.

\noindent ii) {${\bf SL(2,C)}$}: The reduced action (22) is
invariant under the global $SL(2,C)$
transformation. Following Schwarz [10] the part of $S$ coming from the
internal coordinates can be obtained from a $SL(2,C)/SO(2,C)$ coset
construction. We reexpress the action in terms of vielbein fields in
order to show its manifest $SL(2,C)$ invariance.
Let us introduce the $SL(2,C)$ matrix

$$V_j=\bigg (\matrix {{e^{\lambda_jD_j/2}} &
{0}\cr {e^{-\lambda_jD_j/2}a_j}
& {e^{-\lambda_jD_j/2}} \cr }\bigg ). \eqno(32)$$

\noindent Under a global $SL(2,C)$ transformation, $S_j$, and a local
$SO(2,C)$ transformation $O_j$, $V_j\rightarrow O_jV_jS_j$. For a given $S_j$,
an $O_j$ can be always chosen to preserve the form of $Vj$.
So the symmetric matrix,

$$T_j=V^T_jV_j=\bigg (\matrix {{e^{\lambda_jD_j}+e^{-\lambda_jD_j}a_j^2} &
{e^{-\lambda_jD_j}a_j}\cr {e^{-\lambda_jD_j}a_j}
& {e^{-\lambda_jD_j}} \cr }\bigg ), \eqno(33)$$

\noindent transforms as, $T_j\rightarrow S^T_jT_jS_j$. The action can
be rewritten in terms of $T_j$'s as follows,

$$S_4 = \int d^4x \sqrt {g}~\bigg [- R-{1\over 4}g^{rr} \sum_{j=1}^3
\bigg \{{\rm tr}(\partial_rT_j\partial_rT^{-1}_j)\bigg \}\bigg], \eqno (34)$$

\noindent showing that $\lambda_jD_j$ and $a_j$ parametrize the coset
$SL(2,C)/SO(2,C)$. From the transformation law of $T_j$, it is clear
that the reduced action (22) is invariant under the global $SL(2,C)$
transformation. It should be understood that there is an $SL(2,C)$
transformation for each $T_j$. The invariance of the reduced action
under the aforementioned $SL(2,C)$ is due to the specific
decomposition of the background fields adopted in (21).
Consequently we can generate new
backgrounds $\tilde a$ and $\tilde D$ through the implementation of
$SL(2,C)$ trnasformations. For a general $SL(2,C)$ matrix,

$$S_j=\bigg (\matrix {{g_j} &
{d_j}\cr {c_j}
& {b_j} \cr }\bigg ), \qquad g_jb_j-c_jd_j=1,\eqno(35)$$

\noindent where all entries can be complex, the new backgrounds are
given by,

$$e^{-\lambda_j\tilde D_j}={1\over R^2}[{d^2_jR^4_0\over
{Q_j}}+b^2_jQ_j]+ 2ib_jd_jR,' \eqno (36)$$

\noindent and
$$\tilde a_j = {{{{1\over R^2}[{g_jd_jR^4_0\over
{Q_j}}+b_jc_jQ_j] +iR'(d_jc_j+b_jg_j)}}\over {{1\over R^2}[{d^2_jR^4_0\over
{Q_j}}+b^2_jQ_j]+ 2ib_jd_jR'}}. \eqno(37)$$

\noindent In above equations, (36) and (37), we have
used the solution (28) for the old fields.
Now we give two specific examples for illustration. First is a simple
transformation when $g_j=b_j=0$ and $d_j=-c_j=1$. This takes
$T_j\rightarrow T_j^{-1}$. This transformation is same a the target
space duality transformation discussed above.
The second one is more interesting. In this case we choose $S_j$ as
follows,
$$S_j=\bigg (\matrix {{-{{iQ_j}\over {2R^2_0}}} &
{1}\cr {-{1\over 2}}
& {{iR^2_0}\over {Q_j}} \cr }\bigg ), \eqno(38)$$

\noindent and the new fields are given by,
$$e^{-\lambda_j\tilde D_j}={-{{2R^2_0R'}\over {Q_j}}}, \qquad
{\rm and } \qquad \tilde
a_j = {{iQ_j}\over {2R^2R'}}. \eqno(39)$$

\noindent Here we observe that both, radial dilaton and axion become
singular at
$r$=0. Dilaton has constant asymptotic value whereas axion vanishes
at the infinity.

It is natural to ask whether the $SL(2,C)$ symmetry alluded above is
some new symmetry or a part of the $O(d,d)$ symmetry described
earlier. After a little bit of algebra one realises that $SL(2,C)$ is
indeed a part of the $O(d,d)$. The $O(6,6)$ transformation,

$$\Omega = \pmatrix {{\cal B}_1
& 0 & 0 & {\cal D}_1 & 0 & 0\cr 0 &{\cal B}_2
& 0 & 0 & {\cal D}_2 & 0 & \cr
0 & 0 & {\cal B}_3
& 0 & 0 & {\cal D}_3 \cr {\cal C}_1
& 0 & 0 & {\cal G}_1 & 0 & 0\cr 0 &{\cal C}_2
& 0 & 0 & {\cal G}_2 & 0 & \cr
0 & 0 & {\cal C}_3
& 0 & 0 & {\cal G}_3 \cr}, \eqno (40)$$

\noindent with

$${\cal B}_j=\bigg (\matrix {{b_j} &
{0}\cr {0}
& {b_j} \cr }\bigg ), \quad {\cal C}_j=\bigg (\matrix {{0} &
{c_j}\cr {-c_j}
& {0} \cr }\bigg ), \quad {\cal D}_j=\bigg (\matrix {{0} &
{-d_j}\cr {d_j}
& {0} \cr }\bigg ), \quad  {\rm and} {\cal G}_j=\bigg (\matrix {{g_j} &
{0}\cr {0}
& {g_j} \cr }\bigg ),\eqno(41)$$

\noindent induces the same transformation as the
$SL(2,C)$ transformation
given in (35). The $b_j$'s, $c_j$'s etc. used in equation (41) are
the same as those used in eqn. (35).

\noindent {\bf V. Wormholes in string theory: A general case}

In this section we shall give a more general treatment of wormhole
solutions arising from string theory.
We shall discuss how one can get these solutions from various
compactification schemes. Our starting point is the action (11)
and we set $\phi$=constat,
$B_{\mu\nu}$=const. and ${\cal A}_{\mu}^i$=0. The action reduces to,

$$S_ g =C \int d^Dx \sqrt {g}
\bigg\{- R -{1 \over 8} {\rm tr} (\partial_\mu M^{-1} \partial^\mu
M)\bigg\}. \eqno (42)$$

Coming to the equations of motion, we note that two of the equations (18)
and (19) are trivially satisfied. The equations (16), (17) and (20) reduce
to,

$$R_{\mu\nu} -{1\over 2} g_{\mu\nu} R
= {1 \over 16}g_{\mu\nu} {\rm tr} (\partial_\lambda M^{-1} \partial^\lambda
M)- {1 \over 8} {\rm tr} (\partial_\mu M^{-1} \partial_\nu M),
 \eqno(43)$$

$$R +{1 \over 8} {\rm tr} (\partial_\mu M^{-1} \partial^\mu M)=0,
\eqno (44)$$

\noindent and
$$\partial_\mu ({\sqrt g} g^{\mu\nu}M\eta \partial_\nu M)=0.\eqno(45)$$

\noindent Notice, if one contracts eqn. (43) with $g^{\mu\nu}$ one gets the
eqn. (44), and hence eqn. (44) does not give any new information.
Thus, in what follows, we shall work with only equations (43) and (45).
Now, if $M$ depends on $r$ only and space-time metrics are diagonal
and of the form (28a) and (30a), discussed earlier, equations (43) and
(45) will reduce to
much simpler forms given respectively by,

$$R_{rr} -{1\over 2} g_{rr} R
=- {1 \over 16} {\rm tr} (\partial_r M^{-1} \partial_r M), \eqno(46)$$

$$R_{ii}=0,\quad i\neq r,\eqno (47)$$
\noindent and

$$\partial_r({\sqrt {g(r)}}{\sqrt {g(\Theta)}} M\eta \partial_r M)=0.
\eqno(48)$$

In last equation $g(r)$ and $g(\Theta)$ are the $r$ dependent and
the angular ($\Theta$ stands collectively for the rest of the coordinates)
part of the determinant $g$ respectively.
Since, in equation (48) only derivative with respect to $r$ is
appearing, the angular dependence can be factored out
and the equation can be integrated. The result is

$${\sqrt {g(r)}}M\eta \partial_r M=X \eqno (49)$$

\noindent where $X$ is a constant matrix. From its definition (see
Ref. [6] for cosmological solutions) $X$ has following properties.

$$X^T=-X \qquad {\rm and }\qquad M\eta X=-X\eta M. \eqno (50)$$

\noindent Now it is easy to show that

$${\rm tr} (\partial_r M^{-1} \partial_r M)=-{1\over g(r)}{\rm
tr}(X\eta)^2. \eqno(51)$$

\noindent  Using last eqn.(51) in eqn. (46) one has

$$R_{rr} -{1\over 2} g_{rr} R={1 \over {16g(r)}}{\rm tr}(X\eta)^2. \eqno(52)$$

\noindent It can now be shown that eqn. (47)
is consistently  satisfied  for the topologies $R^1\times S^3$ and
$R^1\times S^1\times S^2$. In order to verify these relations, first
one has to calculate
$R_{ii}$ for the specified topology and then has to use the eqn. (52) in that.

Therefore, we have shown that any $M$ satisfying eqns. (49) will produce
wormhole solutions of aforementioned topologies. The scale factor $R(r)$
appearing in the metric will be determined from the equation (52).
The task reduces to finding  $M$ and $X$ which will satisfy eqns.
(49) and (50)
consistently. We now proceed to present the
equations to be satisfied in terms of
$G$ and $B$. In order to fulfill the requirements, we decompose
 the constant matrix $X$ in four
$d\times d$ blocks.

$$X=\bigg (\matrix {{A} &
{C}\cr {-C^T}
& {F} \cr }\bigg ). \eqno(53)$$

\noindent $A^T=-A$ and $F^T=-F$. Notice that $A$, $C$ and $F$ are all constant
matrices. The equation (49) in component form
becomes,

$${\sqrt {g(r)}}(G^{-1}B'G^{-1})=A, \eqno(54)$$
$${\sqrt {g(r)}}(G^{-1}G')-AB=C, \eqno(55)$$
\noindent and
$${\sqrt {g(r)}}(-B'+G'G^{-1}B)+BC=F. \eqno(56)$$

As an illustrative example we reexpress the solution of topology
$R^1\times S^3$ presented in the section IV in
this language. Choosing the ansatz (21) for $G$ and $B$ and using the
solution (28b) we find
consistent $A$, $C$ and $F$ which satisfy eqns. (54)-(56). They are given by,

$$A = \pmatrix {A_1 & 0 & 0 \cr 0 & A_2 & 0 \cr
0 & 0 & A_3 \cr}, \quad C=0, {\rm and} \quad
F = \pmatrix {F_1 & 0 & 0 \cr 0 & F_2 & 0 \cr
0 & 0 & F_3 \cr},\eqno (57)$$

\noindent where $A_j$'s and $F_j$'s are given by,

$$A_j= \pmatrix {0 & 2iQ_j\cr -2iQ_j & 0\cr},\quad{\rm and}\quad
F_j= \pmatrix {0 & -{2iR_O^4\over Q_j}\cr {2iR_O^4\over Q_j}& 0\cr}.
  \eqno (58)$$
\noindent With above form of $X$ and the metric (28a)
 the eqn. (52) reduces to equation (3).
We expect to solve for more $G$ and $B$'s which will satisfy the equations
(54)-(56) and give wormhole solutions.

\noindent {\bf VI. Summary and discussions:}

To summarize our results: We have presented new axionic wormhole
solutions for $R^1\times S^3$ as well as $R^1\times S^1\times S^2$ geometries.
We have shown that the string effective action is invariant under a
global $SL(2,C)$ symmetry for suitable choice of background fields.
Furthermore, it was shown that this $SL(2,C)$ is a part of the
noncompact global $O(d,d)$ symmetry. It is quite
transparent that each wormhole solution is characterized by a
conserved global axionic charge. Thus the new wormhole solutions,
generated through the implementation of the $SL(2,C)$ and duality
transformation, indeed correspond to wormholes with distinct global
charges. Moreover, the radial and axion fields in the solution generated
through the $SL(2,C)$ transformation, have completely different behaviour from
the one they were generated. We have discussed the possibility of
obtaining wormhole solutions in a general setting when the metric
configurations correspond to $R^1\times S^3$ or $R^1\times S^1\times S^2$
geometry. It is interesting to note that the analog of the axionic
charge conservation law, in this case plays a very crucial role in
imposing constraints on the structure of the $M$-matrix which in turn
responsible for the existence of the wormhole solution.

We have mentioned earlier that the wormholes are expected to play
imporatnt roles in quantum gravity at the Planck scale. Hawking has
introduced the concept of vertex operators in order to account for
the effects of wormholes at distances much larger than the
Planck scale. Therefore, it will be interesting to obtain the vertex
operators for the wormhole solutions presented here. It looks quite
promising to adopt the path integral technique envisaged in Ref. [22]
to construct the vertex operators for the wormhole solutions
discussed by us; work is in progress along these directions [28].

The illustrative examples presented here do not exhaust the whole
class of wormhole solutions. It is quite evident that the full
$O(d,d)$ transformations together with the transformations considered
in this in this work will unravel a more rich class of solutions.

\noindent {\bf Acknowledgements:} We are grateful to Ashoke Sen for pointing
out to us that the $SL(2,C)$ symmetry may be a part of the $O(d,d)$
symmetry.

\def \np {{\it Nucl. Phys. }}
\def \pl {{\it Phys. Lett. }}
\def \prl {{\it Phys. Rev. Lett. }}
\def \pr {{\it Phys. Rev. }}
\def \mpl {{\it Mod. Phys. Lett. }}

\noindent {\bf References:}

\item {[1]} S. P. Khastgir and J. Maharana, Inst. of Physics,
Bhubaneswar preprint, IP/BBSR/92-77 (to appear in \pl).

\item {[2]} C. G. Callan, D. Friedan, E. J. Martinec, and M. J. Perry,
{\it Nucl. Phys.} {\bf B262} (1985) 593; A. Sen, {\it Phys. Rev. Lett.}
{\bf 55} (1985) 1846; \pr {\bf D32} (1985) 2102.

\item {[3]} K. Kikkawa and M. Yamasaki, {\it Phys. Lett.} {\bf 149B}
(1984) 357; N. Sakai and I. Senda, {\it Prog. Theor. Phys.} {\bf 75}
(1986) 692; T. Busher, {\it Phys. Lett.} {\bf 194B} (1987) 59;
{\it Phys. Lett.} {\bf 201B} (1988) 466;
{\it Phys. Lett.} {\bf 159B} (1985) 127; V. Nair, A. Shapere,
A. Stominger, and F. Wilczek, {\it Nucl. Phys.} {\bf B287} (1987) 402;
A. Giveon, E. Rabinovici,
and G. Veneziano, {\it Nucl. Phys.} {\bf B322} (1989) 167;
M. J. Duff, \np {\bf B335} (1990) 610.

\item {[4]} A. A. Tseylin and C. Vafa, {\it Nucl. Phys.} {\bf B372}
(1992) 443;
A. A. Tseytlin, {\it Class. Quan. Gravity} {\bf 9} (1992) 979;
A. A. Tseytlin, {\it DAMTP}-15-1992.

\item {[5]} K. A. Meissner and G. Veneziano, {\it Phys. Lett.} {\bf 267B}
(1991) 33; G. Veneziano, {\it Phys. Lett.} {\bf 265B} (1991) 287.

\item {[6]} K. A. Meissner and G. Veneziano, {\it Mod. Phys. Lett.} {\bf A6}
(1991) 3397.

\item {[7]} A. Sen, \pl {\bf 271B} (1991) 295;
A. Sen, \pl {\bf 274B} (1991) 34;
S. F. Hassan and A. Sen, \np {\bf B375} (1992) 103.

\item {[8]}  J. Maharana and J. H. Schwarz, Caltech. preprint,
CALT-68-1790, \np (in press).

\item {[9]} A. Shapere, S. Trivedi, and F. Wilczek, \mpl {\bf A6} (1991) 2677;
A. Sen, Tata preprint, TIFR/TH/92-46 (Sep. 92).

\item {[10]} J. H. Schwarz, Caltech preprint, CALT-68-1815;
J. Maharana and J. H. Schwarz (unpublished).

\item {[11]} E. Witten, {\it Phys. Rev.} {\bf D44} (1991) 314;
R. Dijgraaf, E. Verlinde, and H. Verlinde, \np {\bf B371} (1992) 269;
J. H. Horne and G. T. Horowitz,  {\it Nucl. Phys.} {\bf B368}
(1992) 444; N. Ishibashi, M. Li, and A. R. Steif, {\it Phys. Rev. Lett.}
{\bf 67} (1991) 3336; E. B. Kiritsis, {\it Mod. Phys. Lett.}
{\bf A6} (1991) 2871; S. Kar and A. Kumar, \pl {\bf 291B} (1992) 246,
and references therein; S. F. Hassan and A. Sen, Tata Institute
preprint, TIFR-TH-92-61 (hep-th/9210121), Oct. 92.

\item {[12]} M. Gasperini, J. Maharana, and G. Veneziano,
{\it Phys. Lett.} {\bf 272B} (1991) 167;
M. Gasperini and G.Veneziano, {\it Phys. Lett.} {\bf 277B} (1991) 256;
M. Gasperini, J. Maharana, and G. Veneziano,
{\it Phys. Lett.} {\bf 296B} (1992) 51.

\item {[13]} G. W. Gibbons and K. Maeda, {\it Nucl. Phys.} {\bf B298}
(1988) 741;
C. G. Callan, R. C. Myers, and M. J. Perry, {\it Nucl. Phys.} {\bf B311}
(1988/89) 673; H. J. Vega and N. Sanchez, {\it Nucl. Phys.} {\bf B309}
(1988) 552; {\it Nucl. Phys.} {\bf B309} (1988) 557.

\item {[14]} G. Mandal, A. M. Sengupta, and S. R. Wadia,
{\it Mod. Phys. Lett.}
{\bf A6} (1991) 1685; D. Garfinkle, G. T. Horowitz, and A. Strominger,
{\it Phys. Rev.} {\bf D43} (1991) 3140;
S. P. de Alwis and J. Lykken, {\it Phys. Lett.} {\bf 269B} (1991) 264;
A. Tseytlin,  {\it Mod. Phys. Lett.}
{\bf A6} (1991) 1721; A. Giveon, {\it Mod. Phys. Lett.}
{\bf A6} (1991) 2843; S. Kar, S. P. Khastgir, and A. Kumar,
\mpl {\bf A7} (1992) 1545; A. Sen, \prl {\bf 69} (1992) 1006.

\item {[15]} S. Coleman, \np {\bf B310} (1988) 643;

\item {[16]} E. Baum, \pl {\bf 133B} (1983) 185;
S. W. Hawking, \pl {\bf 134B} (1984) 403.

\item {[17]} S. Coleman, \np {\bf B307} (1988) 867;
S. B. Giddings and A. Strominger, \np {\bf B307} (1988) 854.

\item {[18]} S. W. Hawking, \pl {\bf 195B} (1987) 337; \pr {\bf D37}
(1988) 904;
G. V. Lavrelashvili, V. Rubakov, and P. G. Tinyakov, {\it JETP Lett.}
{\bf 46} (1987) 167; K. Lee \prl {\bf 61} (1988) 263.

\item {[19]} S. B. Giddings and A. Strominger, \np {\bf B306} (1988) 890.

\item {[20]} S. W. Hawking, D. N. Page, and C. N. Pope, \np {\bf B170} (1980)
283; S. W. Hawking, {\it Commun. Math. Phys.} {\bf 87} (1982) 395;
S. W. Hawking, \np {\bf B244} (1984) 135.

\item {[21]} J. Preskill, \np {\bf B323} (1989) 141;
I. Klebanov, L. Susskind, and T. Banks, \np {\bf B317} (1989) 665;
W. Fishler, I. Klebanov, J. Polchinski, and L. Susskind,
\np {\bf B327} (1989) 157; W. Fishler and L. Susskind,
\pl {\bf 217B} (1989) 48; G. W. Gibbons and S. W. Hawking,
\prl {\bf 69} (1992) 1719; {\it Commun. Math. Phys.} {\bf 148} (1982) 345;
B. Grienstein,  \np {\bf B321} (1989) 439.

\item {[22]} B. Grinstein and J. Maharana,  \np {\bf B333} (1990) 160;
S. W. Hawking, \np {\bf B335} (1990) 155;
B. Grinstein, J. Maharana, and D. Sudarsky, \np {\bf B345} (1990) 231.

\item {[23]} B. J. Keay and R. Laflamme, \pr {\bf D40} (1989) 2118.

\item {[24]}  A. K. Gupta, J. Hughes, J. Preskill, and M. Wise,
\np {\bf B333} (1990) 195.

\item {[25]}  J. Scherk and J. H. Schwarz, \np {\bf B153} (1979) 61.

\item {[26]} M. S. Morris and K. S. Thorne {\it Am. J. Phys.} {\bf
56}(5) (1988) 395.

\item {[27]} S. B. Giddings and A. Strominger \pl {\bf 230B} (1989) 46.

\item {[28]} S. P. Khastgir and J. Maharana, (work in progress).

\end